\documentclass[preprint,preprintnumbers,amsmath,amssymb]{revtex4}
\usepackage{amsmath}
\usepackage{txfonts}

\usepackage{graphicx}
\usepackage{rotating}
\usepackage{dcolumn}
\usepackage{bm}

\begin{document}

** CHIN. PHYS. LETT. Vol. 30, No. 8 (2013) 080202

\title{From nothing to something: discrete integrable systems}
\author{S Y Lou$^{1,2}$, Yu-qi Li$^{1,2}$ and Xiao-yan Tang$^{3}$}
\affiliation{$^{1}$Shanghai Key Laboratory of Trustworthy Computing, East China Normal University, Shanghai 200062, China\\
$^{2}$Faculty of Science, Ningbo University, Ningbo,
315211, China\\
$^{3}$Department of Physics, Shanghai Jiao Tong
University, Shanghai, 200240, China}

\date{\today}

\begin{abstract}
Chinese ancient sage Laozi said that everything comes from \emph{\bf \em `nothing'}. \rm Einstein believes the principle of nature is simple. Quantum physics proves that the world is discrete. And computer science takes continuous systems as discrete ones. This report is devoted to deriving a number of discrete models, including well-known integrable systems such as the KdV, KP, Toda, BKP, CKP, and special Viallet equations, from `nothing' via simple principles. It is conjectured that the discrete models generated from nothing may be integrable because they are identities of \emph{\bf \em simple} algebra, model-independent nonlinear superpositions of a \emph{\bf \em trivial} integrable system (Riccati equation), index homogeneous decompositions of the \emph{\bf \em simplest} geometric theorem (the angle bisector theorem), as well as the M\"obious transformation invariants.

\leftline{\bf PACS numbers: \rm 02.30.Ik, 02.90.+p,02.30.Jr\quad DOI: 10.1088/0256-307X/30/8/080202}
\end{abstract}


\maketitle
Around the 6th century BC, in chapter 42 of a Chinese classical text `\emph{Daodejing}', the sage author Laozi said `\emph{Dao sheng yi, yi sheng er, er sheng san, san sheng wanwu, $\cdots$}'\cite{TaoTeJing}. The explanations of traditional philosophers on `Dao sheng yi' is `one comes from nothing' \cite{Zhu}. However, we think that the traditional comprehension on `Dao sheng yi' is not exactly correct. Our understanding on Laozi's original idea is that `Dao' may be a rule, a law, a method, and/or a theory. Using the `Dao', one can produce (`sheng') the first thing (`yi') from nothing. In this sense, we translate Laozi's ideology as `Nothing produces the first via `Dao', then the first produces the second, the second produces the third, and the third produces  everything, ...'. In other words, the essence of Laozi's philosophy is that everything comes from nothing through a suitable `Dao'!

In mathematics, we may write `nothing' as `0' which may also be written as a trivial `identity',
\begin{equation}
0 =x +(-x).\label{nothing}
\end{equation}
Clearly, the Big Bang Theory (BBT) on our universe
and the Dirac sea in quantum physics are two of the best manifestations of the Laozi's philosophy. In the BBT, the universe comes from a singularity point, `0', by a `Big Bang'. Whence the Big Bang occurs, the equal matter `$x$' and antimatter `$-x$' are burst out. In quantum physics, the vacuum, `0', may be considered as a `Dirac sea'. Any particle, `$x$', and its inverse one, the  antiparticle `$-x$', are included in the Dirac sea. When a particle `$x$' is excited from the Dirac sea, its antiparticle `$-x$' will be excited at the same time.

In this Letter, we are interested in whether Laozi's idea can be really used to produce some nontrivial things, discrete integrable models, from `nothing' via a suitable `Dao'.

To see the importance of discrete integrable systems, one may refer to the Scientific Programme Reports of Isaac Newton Institute for Mathematical Science \cite{Newton}.
For example, quantum physics, which has been proved to be correct by extensive physical experiments, tells us that the assumption of a space-time continuum is nolonger adequate at the subatomic scale. In this situation, it has been speculated that a coherent theory of quantum gravity requires an inherently discrete description of the fundamental interactions. To describe these discrete phenomena, we need to develop difference models, in particular, discrete integrable systems.

On the other hand, even in macrophysics, where continuous differential systems are still in the dominant position, discrete integrable systems also play a more important role, due to the rapid development of computer science, which enables most of the continuous  differential systems can be successfully solved numerically where the continuous variables have to be discretized.

One of the most important nonlinear integrable system is the celebrated KdV equation, which was first established for waves on shallow water surfaces and lately be used in almost all physical fields \cite{appl}. It is particularly notable as the prototypical example of exactly solvable models and solitons \cite{KdV}. As in the continuous case, the discrete KdV may have some equivalent versions. One of the elegant versions may be the Schwarzian form,
\begin{equation*}
(u_1-u_2)(u_3-u_4)-(u_2-u_3)(u_1-u_4)=0 \label{KdV}
\end{equation*}
where
\begin{equation*}
u_1=u(m,n)\equiv{u},\ u_2=u(m+1,n)\equiv{\bar{u}},\ u_3=u(m,n+1)\equiv\tilde{u},\ u_4=u(m+1,n+1)\equiv\tilde{\bar{u}}. \label{KdVu}
\end{equation*}

To derive the discrete Schwarzian KdV (dSKdV) from `nothing', the trivial identity \eqref{nothing}, we write $x$ as $|1)|2)|3)|4)$, i.e.,
\begin{eqnarray}
0&=&x+(-x)\nonumber\\
&=&|1)|2)|3)|4)-|1)|2)|3)|4). \label{0}
\end{eqnarray}
Now we introduce some operation rules, `Dao', to Eq. \eqref{0}:\\
\begin{equation}\label{anti}
|1)|2)=\delta |2)|1), \quad \delta^2=1 \qquad \mbox{\rm (symmetric or antisymmetric)},
\end{equation}
\begin{equation}\label{trac}
|1)|2)=(1,2), \qquad\qquad\qquad \mbox{\rm (pairable)}.\qquad\qquad\qquad\qquad
\end{equation}
\begin{equation}\label{comb}
[|1)|2)]|3)=|1)[|2)|3)],\ \  \qquad \mbox{\rm (combinable)}.\qquad\qquad\qquad\quad
\end{equation}

Using the above rules, Eq. \eqref{0} becomes
\begin{eqnarray}
0&=&[|1)|2)][|3)|4)]-[|2)|3)][|1)|4)]\nonumber\\
&=&(1,2)(3,4)-(2,3)(1,4). \label{DKdV}
\end{eqnarray}
It is interesting that Eq. \eqref{DKdV} is just the dSKdV equation \cite{dKdV} if we identify
\begin{eqnarray}
(1,2)=u_1-u_2,\ (3,4)=u_3-u_4,\ \cdots.\label{(1,2)}
\end{eqnarray}
In natural science, it is not a surprise that when one inputs one thing to output another. It will be much more important if one input produces many outputs. Here, we use the same rules (i.e. the same `Dao') to find more interesting significant models from `nothing'.

Let us change $x$ in Eq. \eqref{nothing} as $|1)|2)|3)|4)|5)|6)$. Using the same rules in \eqref{anti}-\eqref{comb} leads to
\begin{eqnarray}
0&=&[|1)|2)][|3)|4)][|5)|6)]-[|2)|3)][|4)|5)][|1)|6)]\nonumber\\
&=&(1,2)(3,4)(5,6)-(2,3)(4,5)(1,6) \label{DKP}
\end{eqnarray}
which is just the well known discrete Schwarzian KP/Toda (dSKP/dST) equation \cite{Schief} with the same corresponding relation \eqref{(1,2)} but ($u\equiv u(m,n,k),\ \bar{u}\equiv u(m+1,n,k),\ \tilde{u}\equiv u(m,n+1,k),\ \hat{u}\equiv u(m,n,k+1)$)
\begin{eqnarray}
&&u_1= \hat{{\bar{u}}},\ u_2=\bar{u},\ u_3= \tilde{\bar{u}},\ u_4= \tilde{u},\ u_5= \hat{\tilde{{u}}},\ u_6=\hat{u}\label{DKPu}
\end{eqnarray}
for the dSKP equation and
\begin{eqnarray}
&&u_1=u,\ u_2=\bar{u},\ u_3=\hat{\bar{u}},\ u_4=\hat{\tilde{\bar{u}}},\ u_5=\hat{\tilde{u}},\ u_6=\tilde{u} \label{DST}
\end{eqnarray}
for the dST equation.

Similar to the continuous KdV equation in (1+1)-dimensions, the continuous KP equation plays the same important role in (2+1)-dimensional physics.

\bf \em Remark 1. \rm It is worth to point out that the dSKP equation is just the Schwarzian form \cite{Q4} of the general Q4 system of the ABS list obtained from the multidimensional consistency \cite{ABS}.

In a same way, the known discrete Schwarzian BKP (dSBKP) equation \cite{Schief} can be derived by replacing `$x$' in Eq. \eqref{nothing} by $|1)|2)|3)|4)|5)|6)|7)|8)$ as follows,
\begin{eqnarray}
0&=&|1)|2)|3)|4)|5)|6)|7)|8)-|1)|2)|3)|4)|5)|6)|7)|8)\nonumber\\
&=&|1)|2)|3)|4)|6)|7)|5)|8)-|1)|4)|2)|3)|5)|6)|7)|8)\nonumber\\
&=&[|1)|2)][|3)|4)][|6)|7)][|5)|8)]-[|1)|4)][|2)|3)][|5)|6)][|7)|8)]\nonumber\\
&=&(1,2)(3,4)(6,7)(5,8)-(1,4)(2,3)(5,6)(7,8)\nonumber\\
&\rightarrow &(w-\tilde{\bar{w}})(\hat{\bar{w}}-\hat{\tilde{w}})
(\tilde{w}-\hat{w})(\bar{w}-\hat{\tilde{\bar{w}}})
-(w-\hat{\tilde{w}})(\tilde{\bar{w}}-\hat{\bar{w}})(\tilde{w}-\bar{w})
(\hat{\tilde{\bar{w}}}-\hat{w})=0. \label{BKP}
\end{eqnarray}

In the same way, by selecting
$$x=|1)|2)|3)|4)|5)|6)|7)|8)|9)|10),$$
or
$$x=|1)|2)|3)|4)|5)|6)|7)|8)|9)|10)|11)|12),$$
in \eqref{nothing}, we may obtain many kinds of unknown equations, say,
\begin{eqnarray}
0&=&|1)|2)|3)|4)|5)|6)|7)|8)|9)|10)-|1)|2)|3)|4)|5)|6)|7)|8)|9)|10)\nonumber\\
&=&[|1)|2)][|3)|4)][|5)|6)][|7)|8)][|9)|10)]
-[|2)|3)][|4)|5)][|6)|7)][|8)|9)][|1)|10)]\nonumber\\
&=&(1,2)(3,4)(5,6)(7,8)(9,10)-(2,3)(4,5)(6,7)(8,9)(1,10). \label{10KP}
\end{eqnarray}
and
\begin{eqnarray}
0&=&|1)|2)|3)|4)|5)|6)|7)|8)|9)|10)|11)|12)
-|1)|2)|3)|4)|5)|6)|7)|8)|9)|10)|11)|12)\nonumber\\
&=&|1)|7)|2)|10)|3)|8)|4)|11)|5)|9)|6)|12)
-|1)|8)|2)|11)|3)|9)|4)|12)|5)|7)|6)|10)\nonumber\\
&=&(1,7)(2,10)(3,8)(4,11)(5,9)(6,12)
-(1,8)(2,11)(3,9)(4,12)(5,7)(6,10). \label{12KP}
\end{eqnarray}

It is interesting that the known discrete Schwarzian CKP (dSCKP) system \cite{Schief} can be produced from \eqref{12KP} by a special reduction $\{|10)\rightarrow |7),\ |11)\rightarrow |8),\ |12)\rightarrow |9) \}$:
\begin{eqnarray}
&&(1,7)(2,10)(3,8)(4,11)(5,9)(6,12)
-(1,8)(2,11)(3,9)(4,12)(5,7)(6,10)=0,\longrightarrow \nonumber\\
&&(1,7)(2,7)(3,8)(4,8)(5,9)(6,9)
-(1,8)(2,8)(3,9)(4,9)(5,7)(6,7)=0,\longrightarrow \label{CKP}\\
&&
(\tilde{\tilde{\bar{v}}}-\bar{v})(\tilde{\bar{\bar{v}}}-\bar{v})
(\hat{\tilde{\tilde{v}}}-\tilde{v})(\hat{\hat{\tilde{v}}}-\tilde{v})
(\hat{\hat{\bar{v}}}-\hat{v})(\hat{\bar{\bar{v}}}-\hat{v})
-(\tilde{\tilde{\bar{v}}}-\tilde{v})(\tilde{\bar{\bar{v}}}-\tilde{v})
(\hat{\tilde{\tilde{v}}}-\hat{v})(\hat{\hat{\tilde{v}}}-\hat{v})
(\hat{\hat{\bar{v}}}-\bar{v})(\hat{\bar{\bar{v}}}-\bar{v})=0.\nonumber
\end{eqnarray}

If the distributable property for the constant multiplication,
$$(a+b)|i)=a|i)+b|i)$$
  is introduced, more discrete integrable systems can be produced from \eqref{nothing}. For instance, if we take $x=(b_1+b_2)|1)|2)|3)|4)$ with arbitrary constants $b_1$ and $b_2$,
we have
\begin{eqnarray}
0&=&b_1|1)|2)|3)|4)+b_2|1)|2)|3)|4)-(b_1+b_2)|1)|2)|3)|4)\nonumber\\
&=&b_1[|2)|3)][|1)|4)]+b_2[|1)|2)][|3)|4)]-(b_1+b_2)\delta[|1)|3)][|2)|4)]\nonumber\\
&=&b_1(2,3)(1,4)+b_2(1,2)(3,4)-(b_1+b_2)\delta(1,3)(2,4). \label{Viallet}
\end{eqnarray}
It is remarkable that Eq. \eqref{Viallet} can be a discrete form of both a special Viallet system or the Bilinear KP/Toda equation. In fact, Eq. \eqref{Viallet} is a variant form of the known special discrete Viallet (SDV) system \cite{Viallet} if we realize it via Eq. \eqref{(1,2)}. However, Eq. \eqref{Viallet} becomes a special form of the discrete bilinear KP/Toda equation \cite{bilinear} if taking
\begin{eqnarray}
(1,2)=f(k_1+1,k_2+1,k_3,k_4), (3,4)=f(k_1,k_2,k_3+1,k_4+1),\cdots.\label{bilinear}
\end{eqnarray}

In fact, by means of the same `Dao', the trivial identity \eqref{nothing}
 can produce a number of nontrivial identities, which may correspond to discrete integrable systems by suitable realizations.

Here we define $A^{o}$ and $A^{e}$ as odd and even pairs of $a_n$ if they are completely paired with odd or even commutations, where
\begin{eqnarray}
a_n\equiv |1)|2) \cdots |2n)\equiv \prod_{i=1}^{2n}|i).\label{a}
\end{eqnarray}

Let $A_{i}^{o},\ i=1,\ 2,\ \cdots,\ N_o$ and $A_{i}^{e},\ i=N_o+1,\ 2,\ \cdots,\ N$ are odd and even pairs of $a_{n}$ for all $i$, $A_{i}^{o,e}\neq A_{j}^{o,e},\ \forall i\neq j,\ i=1,\ 2,\ \cdots, N$ and $N_o$ and $N$ are given nonzero integers,
we have the following algebraic identities
\begin{eqnarray}
0=\sum_{i=1}^{N}b_ia_n =\delta\sum_{i=1}^{N_o}b_i A_{i}^o+\sum_{i=N_o+1}^{N}b_i A_{i}^e,\ \qquad \qquad \left(\sum_{i=1}^{N}b_i=0\right).\label{2n}
\end{eqnarray}
Eq. \eqref{Viallet} is just the special case of Eq. \eqref{2n} for $n=4,\ N_o=1$ and $ N=3$.

\bf \em Remark 2. \rm If $A_{i}^{o,e}$ are odd or even pairs for all $i=1,\ 2,\ \cdots, N=(2n-1)!!$, and $b_i$ are all nonzero, then we have only one nonequivalent identity for every fixed $n$. The richness of the model for fixed $n$ comes from two aspects. (1) Some of $b_i$ may be zeros. (2) The realizations (the correspondence relations between the algebraic elements $(i,j)$ and discrete fields) of a given identity may be not unique, say, the realizations \eqref{(1,2)} and \eqref{bilinear} for Eq. \eqref{Viallet}.

Recently, it is found that the integrability of the differential equations and the differential-difference equations were deeply related to the identities in some different mathematical fields. Then some types of matrix identities were then successfully used to solve integrable systems.

For integrable systems, nonlinear superpositions can also be considered as the identities among some special solutions. By using a nonlinear superposition of an integrable system, one can obtain a new solution from others. Another interesting and important application of a nonlinear superposition of a continuous integrable system is that it usually is a proper integrable discreterization of the original continuous model. Then, the following question is interesting:
\em Is a discrete model a nonlinear superposition of an integrable system? \rm
More specifically, \em is the discrete system \eqref{2n} a nonlinear superposition of a suitable integrable model? \rm

To partially answer the above question, we define the index operation $\&^*$ on $(i,\ j)$ as
\begin{eqnarray}
(c\&^*i, d\&^*j)=cd(i,j) \label{c*}
\end{eqnarray}
for arbitrary $c\equiv c(i)$ and $d\equiv d(j)$.

We call a function $F((i,j), i,j=1, 2, ..., n)$ is index $k$ homogeneous if
\begin{eqnarray}
\left. F((i,j), i=1, 2, ..., n)\right|_{k\rightarrow c\&^*k} =c^{q_k}  F((i,j), i=1, 2, ..., n) \label{ci*}
\end{eqnarray}
with a constant $q_k$ and $k=1,\ 2,\ \cdots, n$.

According to the definitions \eqref{c*} and \eqref{ci*}, we have the following theorem \cite{LLT}.

\bf Theorem 1. \rm \em If $A_{i}^{o,e}$ and their linear combinations are index $k$ homogeneous for all $k=1,\ 2,\ \cdots ,\ n$, then \eqref{ci*} with
\begin{eqnarray} \label{Poly1}
(i,j)=u_i-u_j,
\end{eqnarray}
is a nonlinear superposition of the Riccati equation,
\begin{eqnarray}
\dot u=a+bu+cu^2, \label{rct}
\end{eqnarray}
where the dot above $u$ denotes the derivative with respect to time $t$, $a,\ b$ and $c$ are arbitrary functions of $t$, and $u_i,\ i=1,\ 2,\ \cdots,\ n$ are solutions of \eqref{rct}. \rm

In fact, in Ref. \cite{LLT}, we have proved that all the index homogeneous functions $F((i,j),i,j=1,2,\cdots, n)$ are nonlinear superpositions of \eqref{rct}. The dSKP \eqref{DKP}, dSBKP \eqref{BKP}, dSCKP \eqref{CKP} and SDV \eqref{Viallet} are all special index homogeneous examples and then they are all nonlinear superpositions of the Riccati equation \eqref{rct}.

In continuous case, it is known that an integrable model possesses a Schwarzian form which is invariant under the M\"obious transformation (MT) and can be obtained via Painlev\'e analysis \cite{Painleve}.

Another interesting thing is that in the continuous case, the Darboux transformations (DTs) and B\"acklund transformations (BTs)
are deeply related to the MTs. Actually, to localize  nonlocal symmetries related to DT (and/or BT), one will find that the DT (BT) related symmetries are only the infinitesimal form of the MTs \cite{BT}. Thus, the existence of an MT invariant Schwarzian form is one of the important property for integrable models. In discrete case, there is no general method to find the Schwarzian form of a given known integrable discrete equation. The localization procedure of the nonlocal symmetries related to BTs and DTs provides a potential approach to find Schwarzian forms of discrete integrable systems.

Because the Riccati equation \eqref{rct} is form invariant under MT
\begin{eqnarray}
u\rightarrow \frac{a'+b'u}{c'+d'u},  \quad a'd'\neq b'c',\label{MT}
\end{eqnarray}
it is straightforward to prove the following theorem.\\
\bf Theorem 2. \rm \em
The index homogeneous system \eqref{ci*} with \eqref{Poly1} is invariant under the M\"obious transformation
\begin{eqnarray}
u_i\rightarrow \frac{a_1+b_1u_i}{c_1+d_1u_i}, \ i=1,\ 2,\ \cdots . \label{MT}
\end{eqnarray}
with arbitrary constants $a_1,\ b_1,\ c_1,\ d_1$ and $a_1d_1-b_1c_1\neq 0$. \rm

It is known that the geometric theorems of ancient Greece are closely linked with discrete integrable systems, say, the Menelaus theorem and the Carnot theorem are related to the dSKP and dSCKP equations, respectively \cite{Schief}. In Ref. \cite{LLT}, we have also pointed out that the angle bisector theorem is just the discrete Schwarzian KdV equation. Furthermore, the dSKdV \eqref{DKdV}, dSKP \eqref{DKP}, dSBKP \eqref{BKP}, dSCKP \eqref{CKP} and SDV \eqref{Viallet} and even all the models expressed by \eqref{2n} can all be considered as the special index homogeneous decompositions of the angle bisector theorem. For instance, for the KP equation \eqref{DKP}, we can rewrite it as
\begin{eqnarray}
0&=&(2,3)(4,5)(1,6)-(1,2)(3,4)(5,6) \nonumber\\
&=&(1,6)(2,7)-(1,2)(6,7)
\label{DKP1}
\end{eqnarray}
with the decomposition relation
\begin{eqnarray}
(2,7)=(2,3)(4,5),\
(6,7)=(3,4)(5,6). \label{dc2}
\end{eqnarray}

 In conclusion, a number of discrete systems including the well-known discrete Schwarzian KdV, KP, BKP, CKP and special Viallet equations can be strictly derived from `nothing' which coincides with Laozi's philosophy `Dao sheng yi'. A conjecture is proposed to assume that the discrete systems derived from `nothing' are integrable under some possible meanings and suitable realizations.
These systems possess four elegant properties related to integrability: (i) identities of a same algebra \eqref{anti}-\eqref{comb}; (ii) nonlinear superpositions of the Riccati equation \eqref{rct} which are independent of the arbitrary functions $a,\ b$ and $c$; (iii) index homogeneous decompositions of a same geometric identity, the angle bisector theorem; and (iv) invariants under the MT \eqref{MT}.

In addition to the ancient Chinese philosophy, the geometry theorems of the ancient Greeks are full of magic and mystery. In fact, the solvable problems of simple polygons are deeply related to discrete integrable models. In this study, it is further pointed out that various discrete  systems can be produced from \emph{one} basic geometric theorem, which can be derived from nothing! All the models that come from `nothing' in this study can be embedded in the index homogeneous decompositions of the angle bisector theorem.

Einstein's simplicity principle is one of the most profound belief for natural scientists, especially for theoretical physicists.
The gauge theory in quantum physics and the Big Bang theory in cosmology are two typical successful performance of this belief.
From this study, one can find that this belief has also been deeply and perfectly characterized.
Algebraically speaking, various significant discrete systems  are all identities of a simple algebra. Geometrically speaking, the angle bisector theorem is the simplest geometric identity, while various interesting models are its index homogeneous decompositions. In the differential equation realm, the Riccati equation is the simplest nonlinear integrable equation. This study shows that so many algebraic and  geometric identities are \emph{ model independent } nonlinear superpositions of the Riccati equation. Finally, from the symmetry transformation point of view, the fractional transformation, i.e., MT \eqref{MT}, may be one of the simplest symmetry transformation. While all models described here are MT invariants.

Finally, we return back to Laozi's esoteric and mysterious idea `Dao sheng yi'. In contrast to the  traditional comprehension, we conclude that Laozi's original meaning is that everything can be produced from nothing by using suitable `Dao'. To derive the integrable models from nothing or one thing, the `Dao' we used here are the operation rules of the algebra \eqref{anti}-\eqref{comb}, the model independent nonlinear superposition of the Riccati equation, the index homogeneous decompositions of the bisector theorem, and the M\"obious transformation invariance.

\section*{Acknowledgement}
The authors are in debt to the helpful discussions with Professors Hu X. B., Liu Q. P., Jarmo Hietarinta, Frank Nijhoff, Decio Levi, D. J. Zhang and Y. Chen.
The work was sponsored by the National Natural Science Foundation of
China (Nos. 11175092, 11275123 and 10735030), Shanghai Knowledge Service Platform for Trustworthy Internet of Things (No. ZF1213), and
K. C. Wong Magna Fund in Ningbo
University.

\end{document}